\documentclass[aip,apl,preprint,longbibliography,showpacs,groupedaddress]{revtex4-1}
\usepackage{graphicx}
\usepackage{subfigure}
\usepackage{multirow}
\usepackage{dcolumn}
\usepackage{bm}
\usepackage{color}
\usepackage{natbib}
\newcolumntype{d}[1]{D{.}{.}{1.2}}

\begin{document}

\title{CoIr-carbon complexes with magnetic anisotropies larger than
0.2 eV: a density-functional-theory prediction}

\author{Ruijuan Xiao, Michael D. Kuz'min, Klaus Koepernik, and Manuel Richter} 

\affiliation{IFW Dresden e.V., PO Box 270116, D-01171 Dresden, Germany} 

\date{\today}

\begin{abstract}
We report a density-functional study of the heteronuclear CoIr dimer adsorbed
on benzene or graphene. In either case CoIr prefers an upright position 
above the center of a carbon hexagon with the Co atom next to it. The Ir
atom stays away from the carbon ring and thus preserves its free-atom-like
properties. This results in a very large magnetic anisotropy of more than 
0.2 eV per dimer. So high a value should suffice for long-term data storage
at the temperature of liquid nitrogen.
\end{abstract}


\maketitle

Graphene has recently attracted much attention in both
applied and fundamental science.
Owing to its novel properties, graphene opens new
perspectives for
post-silicon electronics\cite{geim07}, spintronics\cite{karpan07}, chemical
sensor\cite{schedin07}, hydrogen storage\cite{lee10}, touch-screen
panel\cite{bae10}, and further applications\cite{allen10}. Recently, our 
density-functional-theory (DFT) investigations revealed that both benzene and 
graphene are suitable substrates to adsorb some
transition-metal dimers 
(TMD) in a magnetic state of potential interest for
high-density magnetic data storage\cite{our09,our10}.
The hexagonal environment provided by the carbon ring
preserves the electronic structure characteristics of the free
Co dimer, which results in a magnetic anisotropy energy (MAE)
of the order of 100 meV per
dimer{\cite{our09}. A systematic DFT study of a whole series of dimer-benzene complexes
including the 3$d$ and 4$d$ TMD Fe$_2$, Co$_2$, Ni$_2$, Ru$_2$, Rh$_2$, and
Pd$_2$ shows that promisingly
large MAE, stable geometry, and stable magnetic ground state
are also found in Ru$_2$Bz (Bz = C$_6$H$_6$).\cite{our10}
These findings may open a way
to enhance the presently available area density of magnetic recording by 
up to three orders of magnitude. 
Considering that a minimum MAE of 40 $kT$ is required for long-term data 
stability~\cite{charap97},
the reported~\cite{our09,our10} MAE could be sufficient to operate
a magnetic storage device at about 30 K. The aim of the present work
is to seek a possibility to further raise the MAE of TMD complexes
in order to extend the expected operation range to the temperature of
liquid nitrogen.

Spin-orbit interaction in a magnetic state is the primary source of
MAE\cite{brooks40}. Recent DFT calculations found that in the Co group, 
along with the
increasing strength of the spin-orbit coupling parameter,
the free dimer MAE increases from Co$_2$ to
Ir$_2$.\cite{strandberg07,strandberg08,blonski09,fritsch08,seivane07}
Thus, depositing $5d$-TMD like Ir$_2$, on Bz or on
graphene (Gr) might be expected to result in a larger MAE
of the dimer-carbon system than in the related $3d$-TMD case.
Previous studies however tell us that chemical binding
to surfaces frequently reduces the
magnetic moment of $4d$ or $5d$ metals due to their relatively 
small intra-atomic exchange
(Stoner) integrals\cite{cabria02}.
For example,
the calculated ground states of Rh$_2$ or Pd$_2$ adsorbed on Bz 
are non-magnetic.~\cite{our10}
Heterodimers of $3d$ and $5d$ elements would combine a large Stoner
integral ($3d$) with strong spin-orbit coupling ($5d$).

Electronic structure analysis suggests that only a
perpendicular (upright) arrangement of the dimer on Bz or Gr
(local symmetry $C_{6v}$) is helpful to
preserve the large MAE of the dimer\cite{our10}. In such a
configuration, Fig. \ref{fig:CoIrBz_str} (a),
the elemental identity of the atom next to the carbon ring determines the binding
characteristics,
while the atom atop contributes mostly to the magnetic properties of the system.
Therefore, choosing light atoms as base atoms and placing heavy atoms
atop may be a way to utilize the stronger spin-orbit coupling in the latter
without sacrificing the high magnetic moment.
Here, we use the heteronuclear dimer CoIr to
test this idea.
The ground-state structures, binding energies, magnetic moments, 
and magnetic anisotropies of CoIr adsorbed on benzene and graphene were
investigated. For comparison and deeper understanding, these
properties were also studied in Ir$_2$Bz, in Co$_2$Bz,
and in free CoIr, Ir$_2$, and Co$_2$ dimers.

Our DFT calculations were performed with a highly accurate all-electron
full-potential local-orbital scheme (FPLO)~\cite{koepernik99}, release
9.00-34~\cite{fplo09}.
The molecular mode of FPLO with free boundary conditions 
and Fermi temperature broadening of 100 K
was used for free dimer and dimer-Bz systems. 
To simulate the interaction
between the dimer and Gr, a 4$\times$4 supercell (space group P1)
of dimension 9.84$\times$9.84$\times$16 \AA$^3$ was used,
large enough to avoid interactions among dimers in neighboring cells. 
The related k-mesh used for linear-tetrahedron-method integrations
contained 6$\times$6$\times$1 points in the Brillouin zone.
All data reported here were obtained using the 
generalized gradient approximation (GGA) with the PBE96
exchange-correlation functional~\cite{perdew96}.
The valence basis set comprised Co (3$s$,
3$p$, 3$d$, 4$s$, 4$p$, 4$d$, 5$s$), Ir (4$f$, 5$s$, 5$p$, 5$d$, 6$s$,  
6$p$, 6$d$, 7$s$), C (1$s$, 2$s$, 2$p$, 3$s$, 3$p$, 3$d$), and 
H (1$s$, 2$s$, 2$p$) states.
Geometry optimization was carried out in a scalar
relativistic mode with a force convergence threshold of 10$^{-2}$ eV/\AA~.

To find 
the lowest-energy geometries and spin states of the CoIrBz complex, 18 kinds
of initial structures were optimized without any symmetry constraints 
for total spin $S$ = 0, 1, 2, and 3. (See Fig. A1(i)-(vi) and (ix)-(xiv) of 
Ref.~\onlinecite{our10}.
The structures (vii) and (viii) were ignored since there the distance 
between metal atoms is too short to be reasonable. For structures (i), (vi), 
(ix), (xi), (xii), and (xiii), 6 new variants appear 
by interchanging the positions of Co and Ir atoms.)
Initial antiferromagnetic and 
ferrimagnetic states were also considered for $S$ = 0 and 1.

For CoIrGr,
9 initial configurations were considered, denoted by $\parallel_{\rm cc}$,
$\parallel_{\rm mm}$, $\parallel_{\rm tt}$, $\perp_{\rm cCoIr}$, $\perp_{\rm cIrCo}$,
$\perp_{\rm mCoIr}$, $\perp_{\rm mIrCo}$, $\perp_{\rm tCoIr}$ and $\perp_{\rm tIrCo}$.
Here, $\perp$ or $\parallel$ mean that the dimer axis is oriented
perpendicular or parallel to the Gr plane, respectively;
c, m, and t denote that Co and Ir are placed above the center of a carbon
ring, above the midpoint of a C-C bond, or on top of a C atom, respectively.
In the $\perp$
configurations, the atom mentioned first is next to the Gr plane. 
All carbon positions were fixed in the plane with a C-C bond length
of 1.42 \AA.

The MAE of CoIrBz or CoIrGr in the
ground state was computed as the
energy difference, $E^{[100]} - E^{[001]}$, 
between states with magnetization direction perpendicular [0~0~1]
and parallel [1~0~0] to the carbon plane.
The in-plane anisotropy was neglected, [1~0~0] was defined to point
from the center to the corner of the carbon hexagon.
In the case of free dimers, [0~0~1] and [1~0~0] stand for the
directions parallel and perpendicular to the dimer axis.
The MAE was evaluated in the fully 
relativistic mode in which spin-orbit coupling is included in all orders. 
As a matter of experience~\cite{our09}, the MAE values obtained within standard
GGA and by including the orbital polarization
correction\cite{eriksson90} give respectively
a lower and an upper estimate of the expected MAE.

Benzene is a good model system to describe the main
characteristics of dimer adsorption on
graphene\cite{our09, belbruno05}. In Fig. \ref{fig:CoIrBz_str}
we present the ground state and the four lowest-energy isomers of the
CoIrBz complex, obtained by the described optimization.
In the ground state,
the CoIr dimer is bound perpendicularly to benzene with Co
next to the carbon ring ($C_{6v}$ symmetry, Fig. \ref{fig:CoIrBz_str}(a)).
The distance between Co and Ir in the free dimer is 2.13 \AA.
This metal-metal bond length is increased by only 3\% in the
cases of upright adsorption (Fig. \ref{fig:CoIrBz_str}(a), (d)), 
while it is enlarged by about 10\%
in the parallel adsorption modes (Fig. \ref{fig:CoIrBz_str}(b), (c), (e)).
The dimer adsorption 
energy, $E_{ad}$, in the ground state is 1.51 eV, which is comparable to the
$E_{ad}$ value of Co$_2$Bz, 1.47 eV (Tab. I).
We also notice that the calculated
$E_{ad}$ of Ir$_2$Bz, 1.74 eV, is larger than that of
Co$_2$Bz or CoIrBz.
For an upright adsorption with Ir next to Bz (not shown in 
Fig. \ref{fig:CoIrBz_str}), we found a ferrimagnetic state with $E_{ad}$ 
= 0.94 eV and $S$ = 1 at almost the same energy as a
ferromagnetic state with $S$=2.

A detailed understanding of adhesion and isomer energies is complex and
beyond the scope of this paper.
It would require consideration of the transition-metal level positions
in different spin states, their different orbital overlap and their
element-specific Stoner integrals. To give just one example,
we note that the Co contribution to the bonding $\pi$-states
of CoIrBz situated at about -8 eV (Fig. \ref{fig:CoIrBz_orbcomp}) is quite
different in the two 
spin channels. This originates from the spin-splitting of the CoIr-$\pi$ states,
right-hand panel of Fig. \ref{fig:CoIrBz_orbcomp}, 
and the resulting unequal energy differences from the Bz-HOMO level.

The ground-state spin of CoIrBz, $S$ = 2, equals that of the free 
CoIr-dimer. This makes a welcome difference to
Ir$_2$, whose moment is completely quenched when it is adsorbed by benzene.
Using the heteronuclear dimer CoIr instead of Ir$_2$
protects the magnetism of the heavy atom since it stays
atop and preserves its free-atom-like properties to a large extent.
In the ground state of the CoIrBz complex, the
magnetic moments of Co and Ir are about 1.8 $\mu_B$ and 2.3 $\mu_B$,
respectively.
Compared with the free CoIr dimer, the magnetic moment of Co decreases while
that of Ir increases (Tab. I).
Fig. \ref{fig:CoIrBz_orbcomp} shows the energy level scheme and the orbital
composition of each level for the ground states of CoIrBz and of CoIr. 
Like in Co$_2$Bz,~\cite{our10} the
adsorption is realized mainly by forming chemical bonds between carbon and the
atom next to the benzene plane. 
For example, the 3$d_{xy}$, 3$d_{x^2-y^2}$ orbitals of the Co
atom and the LUMO of benzene form the $\delta$ states
at -5.9 eV/-1.7 eV (majority
spin channel) and -4.8 eV/-1.5 eV (minority spin channel).
The $\pi$ states around -8
eV in both spin channels mainly consist of the 3$d_{xz}$,
3$d_{yz}$ orbitals of Co and the HOMO of benzene.
At the Fermi level, a two-fold
degenerate singly occupied $\delta^*$ state
is to 90\% composed of the 5$d_{xy}$, 5$d_{x^2-y^2}$ states of Ir.
It is the splitting of this state by the spin-orbit interaction
that determines the MAE of the whole system.

The value of MAE and the
site-resolved spin and orbital moments $\mu_S$ and $\mu_L$ for
both magnetization orientations are listed in Tab. II.
Data for free CoIr and for CoIrGr (discussed below) are included for comparison.
In all systems, the spin moments are nearly
the same for [0~0~1] and for [1~0~0] orientation, but the anisotropy of the
orbital moments is obvious. 
The Ir atom shows large orbital moments in the case of [0~0~1] orientation
and relatively small values in the [1~0~0] case.
The Co atoms behave differently in the free and in the bound dimers.
In free CoIr, the orbital moment of Co in the [0~0~1]
orientation is around 0.9 $\mu_B$, much larger
than in the [1~0~0] orientation. When bound to benzene, the
orbital moment of Co is quenched.
This comes about through a re-distribution of Co-weight between the $\delta$ and
$\delta^*$ states in the minority spin channel, see Fig. \ref{fig:CoIrBz_orbcomp}.
Interaction with Bz leads to a severe reduction of Co-weight in $\delta^*$ in
favor of the $\delta(\delta)$ and $\pi^*(\delta)$ states.
As a result, the singly occupied $\delta^*$ molecular orbital, carrying a large
ground-state
orbital moment of nearly 2 $\mu_{\rm B}$, is situated almost exclusively
on the Ir atom.

The calculated lower and upper estimates of the MAE in 
CoIrBz are about 250 meV and 290 meV,
respectively. Both of them are larger than either of the values for free CoIr.
The reason is that the MAE is proportional to the orbital moment anisotropy
and grows with the strength of spin-orbit coupling~\cite{our10}
and that bonding to benzene results in a significant
increase of the former on the Ir atom (Tab. II).

Most notably, the {\em lower} estimate of
the MAE of CoIrBz is {\em five times} higher than the respective
estimate for Co$_2$Bz.~\cite{our09}
Because the spin-orbit coupling in Ir is much stronger than in Co, the MAE caused
by the Ir orbital moment anisotropy in CoIrBz is correspondingly much larger than
the purely Co-related MAE in Co$_2$Bz. We also notice that Co$_2$Bz has a 
slightly higher
upper MAE estimate than CoIrBz, which is due to~\cite{our10} the larger Racah parameter 
$B$ of Co (145 meV) than of Ir (94 meV). 

Finally, we investigated
the properties of a CoIr dimer deposited on graphene.
Structural optimizations of all 9 initial configurations
confirm that the CoIr dimer is adsorbed by Gr in an upright position with local
symmetry $C_{6v}$, and that the Co atom prefers to occupy the position
next to the Gr plane.
The calculated dimer adsorption energy for the ground state amounts to 0.63 eV, and the
relative energy of two separate adatoms to a bound dimer,
$[E_{\rm CoGr}+E_{\rm IrGr}-E_{\rm Gr}]-E_{\rm CoIrGr}$, is 3.34 eV.
This clearly indicates that the metal
atoms prefer to aggregate as dimers rather than as separate
adatoms. The Co-Ir bond length and the Co-C bond length are 2.18 \AA~ and 2.25
\AA{}, respectively, similar to the CoIrBz case. A band-structure
analysis (not shown) for the ground state geometry and spin of CoIrGr indicates
that the main contribution to the
states near the Fermi level originates from the 
$5d_{x^2-y^2}$ and $5d_{xy}$ orbitals of Ir. 
Like in CoIrBz, the Ir atom in CoIrGr
shows free-atom-like properties and its orbital moment anisotropy gives rise to
a very large MAE estimated to be between 200 meV and 330 meV.

In summary, the CoIr hetero-dimer is expected to bind with benzene/graphene in perpendicular
orientation above the center of a carbon hexagon with the Ir atom farther away from
carbon than the Co atom.
Due to its remote position, the Ir atom largely preserves its free-atom-like properties.
We predict that this kind of binding does not result in a deterioration
of the magnetic moment of the CoIr dimer and
that the MAE of this structure amounts to 0.2 $\ldots$ 0.3 eV.
These numbers exceed all known experimental values of MAE by more than
one order of magnitude.
Judging by the criterion MAE $>$ 40 $kT$,~\cite{charap97}
CoIrGr should be a promising material for ultrahigh-density magnetic
data storage at liquid-nitrogen temperature.
The demonstrated similarity of our results for CoIrBz and CoIrGr
suggests a possibility to use other substrates, like boron-nitride or
various aromatic carbon-hydrates, with a related chance to achieve
chemical self-assembling.

\clearpage
\begin{figure}
\includegraphics[width=0.5\textwidth]{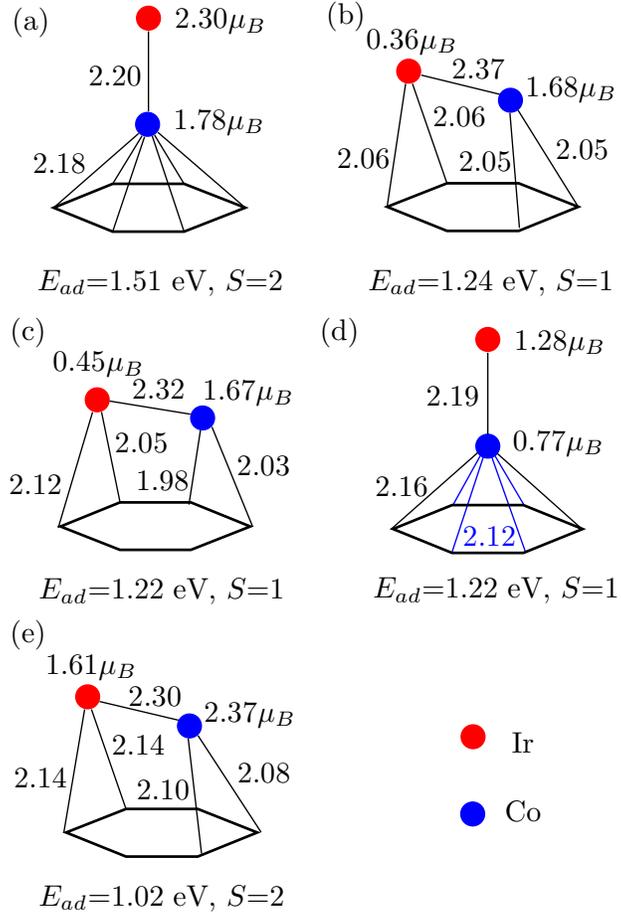}
\caption{\label{fig:CoIrBz_str}(Color online) The dimer adsorption energy ($E_{ad}$), 
metal-metal and metal-carbon bond lengths (in \AA), and spin
magnetic moments for the ground-state geometry (a) 
and higher energy geometries or spin states (b)-(e) of the CoIrBz complex. 
Hexagons indicate benzene rings. The red and blue bullets indicate Ir and Co
atoms, respectively.
}
\end{figure}

\clearpage
\begin{figure}
\includegraphics[width=1.0\textwidth]{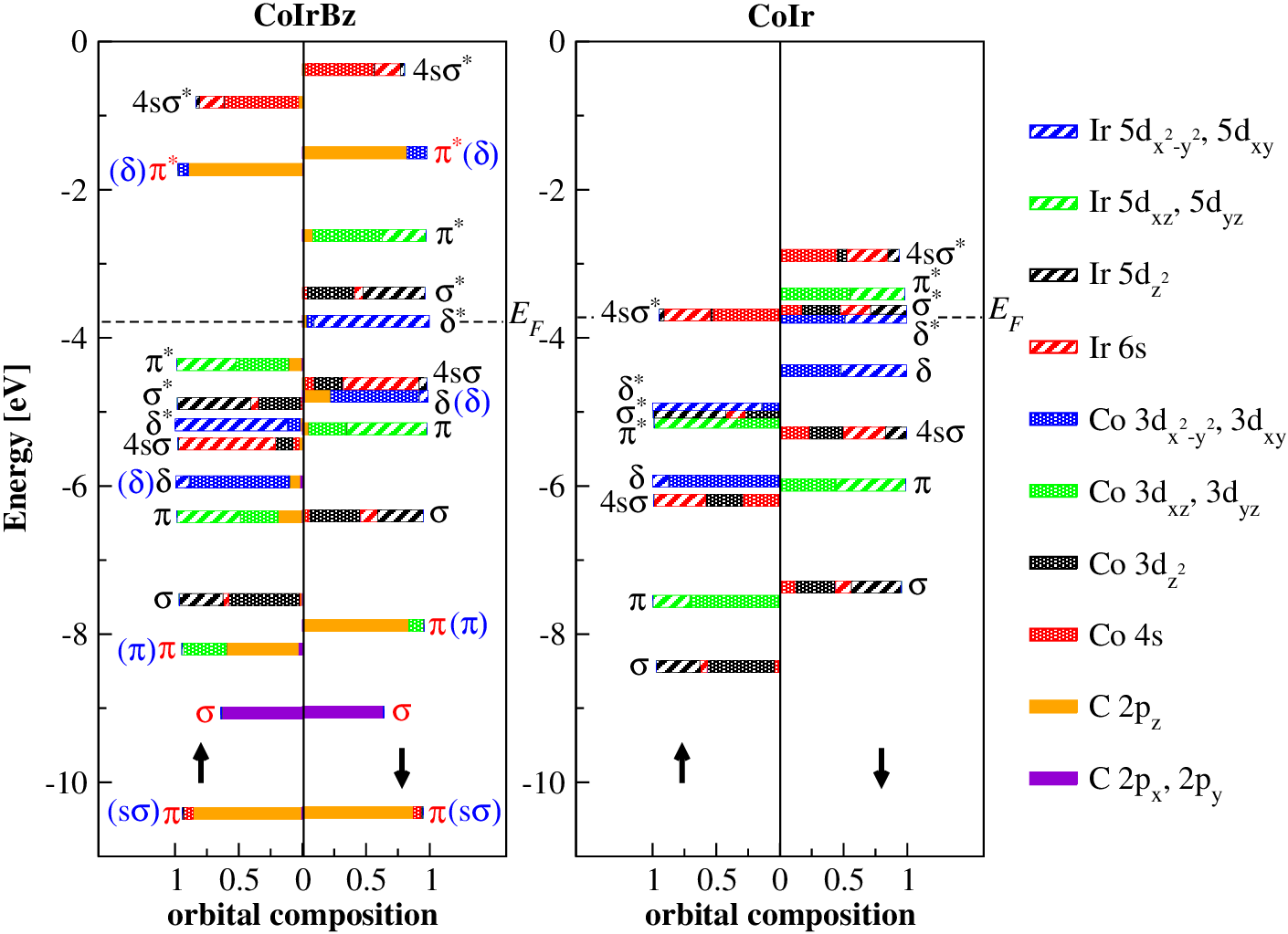}
\caption{\label{fig:CoIrBz_orbcomp}(Color online) Scalar-relativistic
single-particle energy levels and their orbital compositions for the
ground states of CoIrBz (left panel) and CoIr (right panel). The upward
(downward) arrow indicates majority (minority) spin states. All energies refer to
a common vacuum level. Dimer-dominated states are labelled in black and
benzene-dominated states are labelled in red. The three types of chemical bonds
between the CoIr dimer and Bz are labelled blue in parentheses. The position of the
Fermi level ($E_F$) is indicated by a horizontal dashed line. For better resolution,
the $\delta^*_{\uparrow}$-level of CoIr is raised by 0.160 eV,
the $\sigma^*_{\uparrow}$-level is raised by 0.005 eV, and
the $\pi^*_{\uparrow}$-level is lowered by 0.012 eV.
}
\end{figure}

\clearpage
\begin{table}
\caption{Dimer adsorption energy $E_{ad}$, total spin $S$,
atom-resolved spin magnetic
moments, $\mu_{S\rm (TM1)}$ and $\mu_{S\rm (TM2)}$, 
metal-metal and metal-carbon bond lengths, $d_{\rm TM1-TM2}$ and $d_{\rm TM1-C}$, 
for the ground state of Co$_2$,
Co$_2$Bz, Ir$_2$, Ir$_2$Bz, CoIr, CoIrBz, and CoIrGr. TM1 denotes the atom next to
the carbon plane and TM2 the atom farther away.
The data for Co$_2$Bz differ slightly from our previous results~\cite{our09},
because in this work full structural optimization was performed, while
in Ref. \onlinecite{our09} the Bz plane was fixed.
}
\begin{ruledtabular}
\begin{tabular}{cccccccc}
system&Co$_2$&Co$_2$Bz&Ir$_2$&Ir$_2$Bz&CoIr&CoIrBz&CoIrGr\\
\hline
$E_{ad}$ (eV)&$-$&1.47&$-$&1.74&$-$&1.51&0.63\\
$S$ &2&2&2&0&2&2&2\\
$\mu_{S\rm (TM1)}$ ($\mu_B$)&2.00&1.63&2.00&0.00&2.08&1.78&1.81\\
$\mu_{S\rm (TM2)}$ ($\mu_B$)&2.00&2.46&2.00&0.00&1.92&2.30&2.28\\
$d_{\rm TM1-TM2}$ (\AA)&2.00&2.09&2.26&2.25&2.13&2.20&2.18\\
$d_{\rm TM1-C}$ (\AA)&$-$&2.16&$-$&2.28&$-$&2.18&2.25\\
\end{tabular}
\end{ruledtabular}
\end{table}

\clearpage
\begin{table}
\caption{Spin and orbital moments $\mu_S$ and $\mu_L$ (in $\mu_B$)
and MAE ($E_{tot}^{[1 0 0]}-E_{tot}^{[0 0 1]}$, in
meV) of the ground states of CoIr, CoIrBz and CoIrGr.
Values calculated with and without the OP correction are
given in the columns entitled SO+OP and SO, respectively.}
\begin{ruledtabular}
\begin{tabular}{ccccccc}
&\multicolumn{2}{c}{CoIr}&\multicolumn{2}{c}{CoIrBz}&\multicolumn{2}{c}{CoIrGr}\\
\cline{2-3}\cline{4-5}\cline{6-7}
&\multicolumn{1}{c}{SO}&\multicolumn{1}{c}{SO+OP}&\multicolumn{1}{c}{SO}&\multicolumn{1}{c}{SO+OP}&\multicolumn{1}{c}{SO}&\multicolumn{1}{c}{SO+OP}\\
\hline
$\mu_{S\rm (Co)}^{[0 0 1]}$&2.05&2.05&1.74&1.73&1.74&1.73\\
$\mu_{S\rm (Co)}^{[1 0 0]}$&2.14&2.15&1.73&1.74&1.82&1.82\\
$\mu_{L\rm (Co)}^{[0 0 1]}$&0.89&0.95&0.06&$-$0.04&0.09&$-$0.05\\
$\mu_{L\rm (Co)}^{[1 0 0]}$&0.14&0.60&0.23&0.48&0.28&0.65\\
\hline
$\mu_{S\rm (Ir)}^{[0 0 1]}$&1.84&1.82&2.20&2.21&2.17&2.20\\
$\mu_{S\rm (Ir)}^{[1 0 0]}$&1.80&1.84&2.01&2.06&2.04&2.07\\
$\mu_{L\rm (Ir)}^{[0 0 1]}$&1.26&1.23&2.02&2.13&1.83&2.03\\
$\mu_{L\rm (Ir)}^{[1 0 0]}$&0.66&0.96&0.62&0.81&0.63&0.99\\
\hline
$\mu_{L\rm (total)}^{[0 0 1]}$&2.15&2.18&2.08&2.09&1.92&1.98\\
$\mu_{L\rm (total)}^{[1 0 0]}$&0.80&1.56&0.85&1.29&0.91&1.64\\
\hline
MAE&142&230&248&289&198&327\\
\end{tabular}
\end{ruledtabular}
\end{table}

\clearpage

%

\end{document}